\documentclass[
reprint,
superscriptaddress,
 amsmath,
 amssymb,
 aps,
pra,
]{revtex4-2}

\usepackage{graphicx}
\usepackage{dcolumn}
\usepackage{bm}
\usepackage{todonotes} 
\usepackage{physics}
\usepackage{siunitx}
\usepackage[colorlinks=true, linkcolor=blue, citecolor=blue, urlcolor=blue]{hyperref}
\usepackage{siunitx}

\usepackage{braket}

\usepackage{color}
\usepackage{colordvi}  
\definecolor{Red}{rgb}{1,0,0}

\newcommand{\q}{\mathbf{q}}

\newcommand{\tudo}{Condensed Matter Theory, TU Dortmund, Otto-Hahn-Str. 4, 44227 Dortmund, Germany}
\newcommand{\upb}{Institute for Photonic Quantum Systems (PhoQS), Center for Optoelectronics and Photonics Paderborn (CeOPP) and Department of Physics, Paderborn University, 33098 Paderborn, Germany}
\newcommand{\jku}{Institute of Semiconductor and Solid State Physics, Johannes Kepler University Linz, 4040 Linz, Austria}

\makeatother
\makeatletter
\def\authornote2{\xdef\@thefnmark{$\dagger$}\@footnotetext}
\def\authornote3{\xdef\@thefnmark{$\ddagger$}\@footnotetext}
\makeatother
\begin{document}

\title{Experimental measurement of the reappearance of Rabi rotations in semiconductor quantum dots}
\author{Lukas Hanschke$^\dagger$}
\altaffiliation{current address: Walter Schottky Institut, TUM School of Computation, Information and Technology and MCQST, Technical University of Munich, 85748 Garching, Germany}

\affiliation{\upb}

\author{Thomas K. Bracht}
\thanks{These authors contributed equally}
\affiliation{\tudo}
\author{Eva Schöll}
\altaffiliation{current address: Institute of Semiconductor and Solid State Physics, Johannes Kepler University Linz, 4040 Linz, Austria}
\affiliation{\upb}

\author{David Bauch}
\affiliation{\upb}

\author{Eva Berger}
\affiliation{\upb}
\author{Patricia Kallert}
\affiliation{\upb}

\author{Melina Peter}
\affiliation{\jku}
\author{Ailton J. Garcia Jr.}
\affiliation{\jku}
\author{Saimon F. Covre da Silva}
\altaffiliation{current address: Instituto de Física Gleb Wataghin, Universidade Estadual de Campinas (UNICAMP), Campinas, Brazil}
\affiliation{\jku}
\author{Santanu Manna}
\altaffiliation{current address: Indian Institute of Technology, Delhi, New Dehli, India}
\affiliation{\jku}

\author{Armando Rastelli}
\affiliation{\jku}
\author{Stefan Schumacher}
\affiliation{\upb}
\author{Doris E. Reiter}
\affiliation{\tudo}
\author{Klaus D. Jöns}
\affiliation{\upb}
\date{\today}

\begin{abstract}
Phonons in solid-state quantum emitters play a crucial role in their performance as photon sources in quantum technology. For resonant driving, phonons dampen the Rabi oscillations resulting in reduced preparation fidelities. The phonon spectral density, which quantifies the strength of the carrier-phonon interaction, is non-monotonous as a function of energy. As one of the most prominent consequences, this leads to the reappearance of Rabi rotations for increasing pulse power, which was theoretically predicted in Phys. Rev. Lett. 98, 227403 (2007). In this paper we present the experimental demonstration of the reappearance of Rabi rotations.
\end{abstract}

\maketitle
\section{Introduction}

Solid-state quantum emitters offer several advantages for quantum technologies due to their tuneability, robustness and the possibility of integration  \cite{senellart2017,rodt2020deterministically}. In contrast to atoms, solid-state quantum emitters are prone to the interaction with the quantized lattice vibrations, i.e., the phonons. Among the different realizations, semiconductor quantum dots are one of the most promising photon sources \cite{hepp2019semiconductor,zhou2023epitaxial,heindel2023quantum}. The influence of phonons is prominently seen in the phonon sidebands \cite{Besombes2001acoustic,Krummheuer2002theory,favero2003} and has great impact on the state preparation of quantum dots \cite{luker2019review}. Most of the time, phonons hinder a perfect preparation \cite{machnikowski2004resonant, krugel2005role,ahn2005resonance, ramsay2010phonon, mccutcheon2011gen,schuh2011rapid}, but a phonon-assisted process can also lead to state preparation \cite{ardelt2014dissipative,quilter2015phononassisted,bounouar2015phononassisted}. Accordingly, the phonons impact the properties of the generated photons from quantum dots \cite{thomas2021bright,cosacchi2021accuracy} including quantum-dot--cavity systems \cite{roy2011influence,roy2015quantum, sirkina2023impact, bauch2024}. Understanding the electron-phonon interaction is crucial to realize solid-state quantum emitters. 
\begin{figure*}[ht]
	\includegraphics[width=\textwidth]{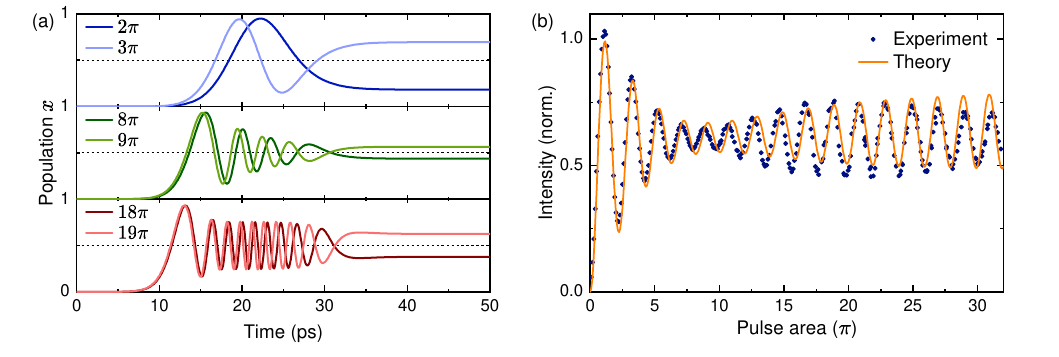}
	\caption{(a) Calculated Rabi oscillations for different pulse areas, varied by adjusting the pulse intensity. The  lower damping of the population evolution for small (blue lines) and large (red lines) pulse areas results in a higher contrast of the final state population compared to intermediate pulse areas (green lines). (b) Rabi rotation obtained for a pulse duration of $\tau = \SI{9.0}{\pico\second}$ at $T = \SI{6.1}{\kelvin}$. The damping of the oscillation reaches its maximum for a pulse area of $\sim 9 \pi$. Further increasing the pulse area leads to a clear reappearance of the Rabi rotation. The theoretical model is in excellent agreement with the experimental data. The experimental data is shifted to the average of the oscillation to emphasize the envelope.}
	\label{fig:hero}
\end{figure*}
The influence of phonons is intricate because of the pure-dephasing-type Hamiltonian that arises due to the discrete energy structure of the quantum dots \cite{mahan2000many}. A well-established consequence of the pure-dephasing Hamiltonian can be seen for an optically driven two-level quantum dot. For resonant driving, Rabi oscillations occur, that become faster with increasing pulse strength (or pulse area). Due to the interaction with longitudinal acoustic (LA) phonons, each oscillation is damped in time, as shown in Fig.~\ref{fig:hero}(a).\\
There, we consider three pairs of pulse areas that differ by $\pi$. Without phonons, for odd integers of $\pi$, full inversion is reached, while for even integers of $\pi$ the occupation after the pulse is zero. With phonons, this contrast of one is reduced, but depends on the pulse area. For small pulse areas [cf. $2\pi/3\pi$], the contrast between the final occupations remains very large. As the pulse area increases [cf. $8\pi/9\pi$], the damping becomes stronger such that the final occupations are both close to $0.5$. With even larger pulse areas [cf. $18\pi/19\pi$], the contrast between the final occupation increases again. This behavior is summarized in Fig.~\ref{fig:hero}(b), which shows the photon emission as a function of pulse area, commonly referred to as Rabi rotations. A non-monotonic behavior of the Rabi rotations as a function of pulse area is observed, with the strongest damping around $9\pi$. This is counter-intuitive to the behavior of Rabi oscillations as a function of time, which are always dampened.\\
The phenomenon of the non-monotonous damping as a function of pulse area was coined Rabi reappearance and was first predicted in 2007 \cite{vagov2007nonmonotonic}. The Rabi reappearance is a direct consequence of the non-monotonic behavior of the phonon coupling strength as a function of energy.  While the underlying theory has been well studied, as summarized in several reviews \cite{grosse2008phonons,carmele2019non, reiter2019distinctive}, the experimental verification remains challenging. Rabi rotations with clear attribution to phonon-damping have been measured \cite{ramsay2010phonon}, but the reappearance regime has not been observed so far.\\ 
In this paper, we show experimental measurements of the reappearance of Rabi rotations in good agreement with theory in Fig.~\ref{fig:hero}(b). The intensity as a function of pulse area shows a clear signature of the Rabi reappearance: Up to about $9\pi$, the amplitude of the Rabi rotations decreases, but above $9\pi$ the amplitude of the Rabi rotations begins to increase again.\\
In the following we provide insights into the physics of the reappearance, followed by details on how we successfully measured the reappearance regime. As a proof, we show measurements for different settings of the laser and sample temperature and analyze the response.

\section{Physics of Rabi Reappearance}
\begin{figure}[h]
	\includegraphics[width=\columnwidth]{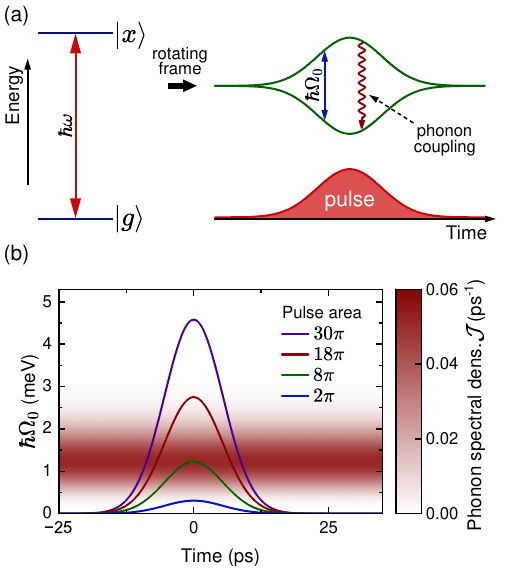}
	\caption{(a) Energy level scheme of a two-level system consisting of the ground $\ket{g}$ and excited state $\ket{x}$ coupled to a resonant laser field. In the rotating frame, dressed states are formed. Their degeneracy is lifted and the splitting $\hbar\Omega_0$ is proportional to the amplitude of the laser pulse as indicated in the bottom. Between the dressed states, phonons can induce transitions, but for low temperatures only phonon emission is allowed. 
    (b) Instantaneous Rabi energy as a function of time. The colored background shows the phonon spectral density $J(\omega)$. When the Rabi frequency and the phonon spectral density agree, an efficient coupling to the phonons takes place. 
    }
	\label{fig:intro}
\end{figure}   

The coupling of phonons to the electronic states in a quantum dot does not lead to a transition between excited $|x\rangle$ and ground state $|g\rangle$ due to the strong energetic mismatch. Instead, the phonons couple only diagonally to electronic states, which is described by a \emph{pure-dephasing} Hamiltonian that accounts for the loss of coherence without energy exchange~\cite{mahan2000many}. 

A clear physical picture of the Rabi rotations can be given in the rotating frame of the resonant laser. There, the two states become degenerate for a vanishing driving strength. In the presence of the laser field this degeneracy is lifted and the resulting dressed states split by the Rabi energy $\hbar\Omega_0$ as depicted in Fig.~\ref{fig:intro}(a). LA phonons now lead to transitions between the dressed states with phonon emission processes from the upper to the lower branch and phonon absorption processes for the reverse path. These processes disturb the coherence of the driven system resulting in damping of the Rabi oscillation. The interaction mechanism to the LA phonon is given by the deformation potential. Due to the nanostructuring, the bulk coupling matrix element of the deformation potential is modified by the confinement of the electronic wave function to the quantum dot via the form factor \cite{luker_phonon_2017}. This leads to the non-monotonic behavior of the phonon coupling strength, that is described by the phonon spectral density $\mathcal{J}$. The bulk deformation potential results in the superohmic coupling and the form factor suppresses the coupling for higher energies. This yields a narrow distribution with a single maximum for the phonon spectral density. This behavior is depicted in Fig.~\ref{fig:intro}(b) as red shaded area. While the exact phonon spectral density depends on the quantum dot geometry \cite{luker_phonon_2017}, it is sufficient to consider a spherical potential. For our chosen parameters (see Appendix A), the phonon spectral density has a maximum around $\hbar\Omega_0\approx 1.2$~meV.

This implies that for constant optical driving, the phonon coupling is resonant when driven with a Rabi frequency corresponding to the maximum of the phonon spectral density at $\hbar\Omega_0\approx 1.2$~meV, where the damping of the Rabi oscillations is maximal. The phonon relaxation time between the dressed states is proportional to the phonon spectral density. However, for pulsed driving, one has to consider the instantaneous Rabi frequency for each moment in time, as given by the splitting between the dressed states. To estimate the phonon coupling, the instantaneous Rabi frequency has to be compared to the phonon spectral density, as displayed in Fig.~\ref{fig:intro}(b). For a pulse area of $2\pi$, the Rabi frequency stays well below the maximum of the phonon spectral density. Thus, an overall weak coupling to phonons is found. Increasing the pulse area to 8$\pi$ pushes the Rabi energy higher, such that it now overlaps with the maximum of the phonon spectral density. This results in strong damping of the Rabi oscillations. A further increase of the pulse area up to $18\pi$ and eventually $30\pi$ sets the Rabi frequency at its maximum far above the maximum of the phonon coupling strength. At this time the phonons become decoupled and do not influence the Rabi oscillations. Following the damping behavior as a function of increasing pulse area yields the reappearance of Rabi rotations.

However, it is important to consider that phonon relaxation processes are possible during the whole time of the pulse and not only its maximum. Accordingly, phonon emission takes place also at the flanks of the pulse. As a result, for example for the $30\pi$ pulse, phonon damping still occurs before and after the pulse maximum, such that the final occupation after the pulse does not reach unity anymore. This can be circumvented by using rectangular pulses that do not allow for an interaction time. Indeed, in this case the final occupation is predicted to reach full population inversion in the reappearance regime \cite{vagov2007nonmonotonic}.

In other optical excitation schemes, the non-monotonic phonon coupling is also visible: For phonon-assisted state preparation schemes relying on the phonon-mediated transition between the dressed states \cite{ardelt2014dissipative,quilter2015phononassisted,bounouar2015phononassisted}, the window of possible detunings is limited. In excitations with chirped laser pulses, resulting in the adiabatic rapid passage effect (ARP), due to the finite splitting between the dressed states, an excitation beyond the phonon spectral density can be achieved more easily \cite{kaldewey2017demonstrating,ramachandran2020suppression}. 

\section{Experimental prerequisites}
 
\begin{figure}[ht]
	\includegraphics[width=\columnwidth]{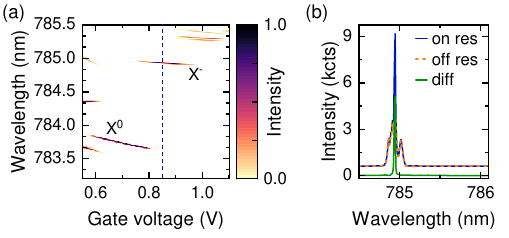}
	\caption{(a) Gate voltage dependent photoluminescence measurement for non-resonant excitation exhibits distinct charge plateaus of the neutral X$^0$ and negatively charged X$^-$ exciton. The blue dashed line indicates the applied gate voltage $V = \SI{0.85}{\volt}$ chosen for resonant excitation. (b) The gated quantum dot structure facilitates the acquisition of spectra under identical conditions with the trion transition tuned in (blue) and out of resonance (dashed orange) with the excitation laser energy. This allows the subtraction of remaining unsuppressed laser to clear the spectrum from unwanted background laser signal (green).}
	\label{fig:gate}
\end{figure}

To observe the reappearance of Rabi rotations experimentally we resonantly drive the negative trion transition of a single droplet etched GaAs quantum dot \cite{dasilva2021gaas} with a few picosecond long laser pulse. The quantum dots are epitaxially grown in the intrinsic region of a p-i-n diode structure close to the n-contact to enable deterministic charging as is shown by gate voltage dependent measurements under non-resonant excitation in Fig.~\ref{fig:gate}(a). In addition, charge noise as an additional source of decoherence is strongly reduced leading to a near Fourier-limited linewidth of the investigated transition. The trion transition provides a clean two-level system due to the intrinsic absence of a fine structure splitting and a far detuned charged biexciton.

To allow a study of resonance fluorescence even under strong driving, the excitation laser must be efficiently filtered from the detection channel. This is realized in a confocal configuration by cross-polarized filtering combined with spatial filtering. This implies a mechanically stable setup as well as a clean flat sample surface to avoid distortion of the beam profile. Further improvement is achieved by placing the quantum dot in a weak planar cavity to increase the coupling of the pulse to the quantum dot as well as the collection efficiency of the emitted photons. Finally, the deterministic charging via the diode structure is utilized to detune the transition far from the laser energy after acquisition of the emission under resonant excitation. This allows recording of an identical spectrum of the remaining laser signal which is then subtracted from the resonant spectrum to isolate the quantum dot emission. An example is depicted in Fig.~\ref{fig:gate}(b). Employing this techniques facilitates recording the Rabi rotation of the negative trion transition by measuring the emission while increasing the laser power and maintaining the pulse duration. Further details on the setup and sample can be found in Appendix B.
        
\section{Measurements of Rabi Reappearance}

Let us get back to the measured data in Fig.~\ref{fig:hero}. In our experimental setup, we observe clean Rabi rotations up to $32\pi$, with maximum damping occuring at around $9\pi$. The temperature was set to $T=\SI{6.1}{\kelvin}$. The experimental data is symmetrically offset to the average of the oscillation to balance for the influence of small detunings of the laser, the temporal pulse shape and further experimental imperfections. The impact of those deviations from the ideal setting are discussed in depth in the Appendix. 

The reappearance of Rabi rotations is clearly visible in the experimental data. To verify this observation, we performed theoretical calculations within the standard model consisting of a two-level system coupled to LA phonons \cite{reiter2019distinctive}. For details of the model we refer to App. \ref{sec:app:theory}. We solve the equations of motion within a numerically exact process-tensor matrix product operator (PT-MPO) algorithm \cite{cygorek2022simulation,cygorek2021code}. The algorithm accounts for all polaron and non-Markovian effects. The only free fitting parameter in the theoretical model is the phonon spectral density. Assuming a spherical dot, which is sufficient for the electronic properties \cite{luker_phonon_2017}, the quantum dot size determines the shape of the phonon spectral density. We furthermore use a slightly enhanced coupling strength to match the data, which is reasonable within our simple model of the phonon spectral density. While varying the temperature, the best agreement was found for a slightly higher temperature of $\SI{8}{\kelvin}$ as in experiment. 

Our theory fits very well to the experimental data. Only at the strongest driving, we find minor derivations, which we attribute to co-tunneling of the trapped charge carriers to the n-doped layer introducing an additional decoherence mechanism and the impact of the carved laser pulse (see Appendix C).

\begin{figure}[h]
	\includegraphics[width=\columnwidth]{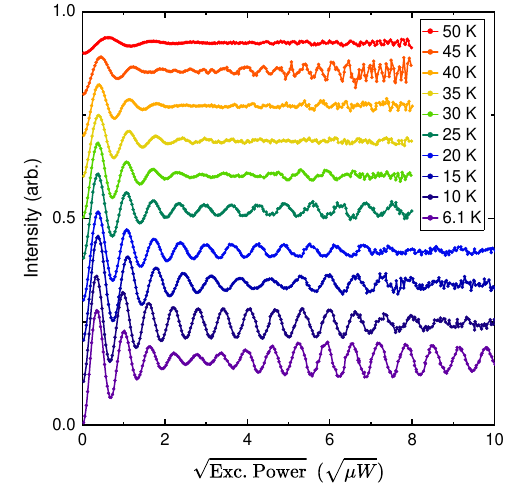}
	\caption{Rabi rotations of the negative trion transition driven by a $\SI{9.0}{\pico\second}$ long laser pulse for varying sample temperature from $\SI{6.1}{\kelvin}$ to $\SI{50}{\kelvin}$. The single data sets are offset along the y-axis for more clarity. Clear indication of the reappearance of the Rabi rotation is even found for elevated temperatures.}
	\label{fig:temp_exp}
\end{figure} 

A temperature series of the Rabi rotations is shown in Fig.~\ref{fig:temp_exp} starting from $6.1$~K up to $50$~K. The individual data sets are offset vertically for clarity. With increasing temperatures the effective phonon coupling increases as more phonons become available for scattering. Accordingly, a stronger damping is expected. Nonetheless, in the strong driving regime, where the phonons are effectively decoupled, the reappearance should be still visible, but less pronounced \cite{vagov2007nonmonotonic}. Indeed, we find in Fig.~\ref{fig:temp_exp} that the damping in particular of the first Rabi rotations becomes stronger. For temperatures up to about $30$~K the reappearance of Rabi rotations is clearly visible, and even for higher temperature we find that the signal becomes stronger again at high excitation powers. 

Also for the temperature series, we performed a theoretical analysis of the data. While we find matching conditions for each temperature individually, when fitting all data simultaneously an adjustment of the phonon coupling strength had to be performed as discussed in App.~\ref{sec:app:temp}. We assume that slight variations in the phonon spectral density occur due to adjustment of the resonance condition via a voltage change, that might result in modification of the wave function.

In addition to the temperature series, we have performed pulse duration and detuning dependent measurements to support the observation of the reappearance of Rabi rotations. The results are discussed in detail in the appendix. We find that the reappearance of Rabi rotations occurs for varying pulse durations from $3$ to $12$~ps with varying minima. Because we vary the pulse duration via a pulse carving process, deviation from the Gaussian shape can occur that lead to additional influences on the Rabi rotations.

\section{Conclusions}
We have experimentally demonstrated the reappearance of Rabi rotations in an optically driven quantum dot under the influence of phonons. The reappearance of Rabi rotations is a direct consequence of the non-monotonic behaviour of the phonon spectral density, which is typical for localized solid-state quantum emitters. However, so far the experimental demonstration has been rather elusive, as it requires certain excitation conditions like strong pulses and stability over a wide range of parameters. 

The measurement was achieved by recent improvements in the sample quality, enabling near Fourier-limited linewidth and thus a strong reduction of other dephasing channels \cite{zhai2020low}, as well as exploiting the electrical switching of the quantum dot-laser coupling to extract the emission spectrum even for strong driving. This made it possible to observe the reappearance of Rabi rotations experimentally for a wide range of laser pulse parameters and sample temperatures. All our results are in excellent agreement with theory and support our finding that we indeed observe the reappearance of Rabi rotations. 

The fundamental insights into the electron-phonon interaction can be transferred from quantum dots to many solid-state quantum emitters \cite{mitryakhin2024engineering}. Understanding electron-phonon interaction is crucial in tailoring their usage for quantum technologies.

\section{Acknowledgements}
We thank Florian Kappe and Gregor Weihs for fruitful discussions on the impact of pulse carving, Moritz Cygorek for his valuable contributions regarding the numerical implementation of the path-integral method and Friedrich Sbresny for his helpful insights into the experimental implementation. 
This work is supported by the Deutsche Forschungsgemeinschaft (German Research Foundation) through the transregional collaborative research center TRR142/3-2022 (231447078, project C09), the ERC grant (LiNQs, 101042672), the Photonic Quantum Computing initiative (PhoQC) of the state ministry (Ministerium für Kultur und Wissenschaft des Landes Nordrhein-Westfalen), and with computing time provided by the Paderborn Center for Parallel Computing, PC$^2$.

\bibliography{thomas}

\newpage

\appendix

\begin{widetext}

\section{Theoretical Model} \label{sec:app:theory}
To model Rabi oscillations in a semiconductor quantum dot, we use a two-level system driven by an external laser pulse. The ground- and excited state $\ket{g}$ and $\ket{x}$ are separated by the energy $\hbar\omega_0$. The laser pulse is included using the time-dependent term $\Omega(t)$ in rotating wave approximation (RWA). In the dipole approximation, $\Omega(t)$ is proportional to the electric field $E(t)$ of the laser and the dipole moment $d$ according to $\Omega(t) \sim d E(t)$. In total, the Hamiltonian for this system reads
\begin{equation}
    H = \hbar\omega_0\ket{x}\bra{x} - \frac{\hbar}{2}\left(\Omega^*(t)\ket{g}\bra{x} + \Omega(t)\ket{x}\bra{g}\right).
\end{equation}
In the RWA, the time dependence of the laser is split into the envelope and a term rotating with the laser frequency $\omega_L$. For a Gaussian shape laser pulse with pulse area $\alpha$, this leads to
\begin{equation}
    \Omega(t) = \Omega_0(t) e^{-i\omega_L t} = \frac{\alpha}{\sqrt{2\pi\sigma^2}} e^{-t^2/(2\sigma^2)}e^{-i\omega_L t}
\end{equation}
While in the experiment the pulse duration is measured by the full-width of half maximum (FWHM) of the autocorrelation function of the intensity of the laser pulse $\tau_{\text{FWHM}}^{\text{corr}}$, in the theoretical model $\sigma$ is a measure for the pulse duration of the electric field. For Gaussian pulses, these are connected by
\begin{equation}
    \tau_{\text{FWHM}}^{\text{corr}} = 2\sqrt{2\ln{2}}\sigma.
\end{equation}
The pulse durations given in the main part of the paper are already divided by $\sqrt{2}$, so for a pulse with $\tau=\SI{9}{ps}$, we use $\sigma=\SI{5.4}{ps}$ for the theoretical calculations. \\
We include a radiative decay rate $\gamma$ of the two-level system using Lindblad operators
\begin{equation}
    \mathcal{L}_{\hat{O},\gamma}\rho = \frac{\gamma}{2}\left(2\hat{O}\rho\hat{O}^\dagger - \hat{O}^\dagger\hat{O}\rho-\rho\hat{O}^\dagger\hat{O}\right).
\end{equation}
For the decay of the exciton state, we use $\mathcal{L}_{\ket{g}\bra{x},\gamma}$. For each pulse area, the time dynamics are calculated until the excited state population decays close to zero, such that the counts $P_{\ket{x}}$ can be calculated using
\begin{equation}
    P_{\ket{x}} = \gamma\int_{-\infty}^{\infty} dt  \braket{\ket{x}\bra{x}}(t).
\end{equation}
We also include the effects of the phonon environment. For this, the two-level system is coupled to longitudinal acoustic phonons using the standard pure-dephasing Hamiltonian reading 
\begin{align}
\begin{split}
        H_{\text{ph}} &= \hbar\sum_{\q}\omega_{\q}^{}b^{\dagger}_{\q}b_{\q}^{}
        + \hbar\ket{x}\bra{x}\lambda\sum_{\q}\left(g_{\q}^{}b_{\q}^{}+g_{\q}^{*}b_{\q}^{\dagger}\right).
\end{split}
\label{eq:hamiltonian_phonons}
\end{align}
The bosonic operators $b^{}_{\q}(b^{\dagger}_{\q})$ describe the destruction (creation) of a phonon with energy $\hbar\omega_{\q}$ and wave vector ${\q}$. We assume a linear dispersion for the phonons and consider deformation potential coupling via the coupling matrix element $g_{\q}$. By adding a factor $\lambda$ to the coupling we scale the phonon coupling for better agreement with experiment. 

The size of the quantum dot dictates the shape of the phonon spectral density $J(\omega)$ \cite{luker_phonon_2017}. We assume a spherical quantum dot, for which the phonon spectral density is 
\begin{equation}
    J(\omega) =  \sum_{\q} |\lambda|^2 |g_{\q}|^2\delta(\omega-\omega_{\q}) = \frac{ |\lambda|^2 \omega^3}{4\pi^2\rho\hbar c_{LA}^5} \left(D_e e^{-\left(\frac{\omega a_e}{2 c_{LA}}\right)}-D_h e^{-\left(\frac{\omega a_h}{2 c_{LA}}\right)}\right)
\end{equation}
We use the parameters for a spherical quantum dot in a GaAS matrix given in Tab.~\ref{tab:phonon_parameters}, in our case the data is best described using a QD size of $\SI{5}{\nm}$.

We note that for the optical excitation, the resonance condition accounts for the polaron shift. For all calculations, initially the QD is in a product state of the ground state $|g\rangle$ and a phonon state following a Bose distribution for temperature $T$ (with $\beta=k_B T$)
\begin{equation}
    n_{\mathrm{ph}}(\q,t=0)=\frac{|g_{\q}|^2}{e^{\beta \hbar \omega_{\mathbf{q}}}-1} \,.
\end{equation}
For the simulation of the dynamics of the system, we employ a state-of-the-art PT-MPO technique detailed in Refs.~\cite{cygorek2022simulation,cygorek2024sublinear,cygorek2024treelike} that allows for a numerically exact description of the system including its interaction with the phonon environment, meaning it makes no approximations other than the temporal discretization. In particular, this method is capable of resolving small sub-picosecond time steps while still capturing the whole non-Markovian nature of the phonons with memory times on the order of several picoseconds \cite{cygorek2024sublinear}.  

\begin{table}[]
    \centering
    \begin{tabular}{l l l}
    Parameter& & Value\\
    \hline
       material density   & $\rho$ & \SI{5370}{\kilo\gram\per\cubic\meter} \\
        sound velocity & $c_{LA}$ & \SI{5.11}{\nano\meter\per\pico\second} \\
       electron deformation potential & $D_e$ & \SI{7}{eV} \\
       hole deformation potential & $D_h$ & \SI{-3.5}{eV} \\
       dot size/electron radius & $a_e$ & \SI{5}{\nano\meter} \\
       hole radius & $a_h$ & $a_e/1.15$ \\
       phonon coupling & $\lambda$ & 1.5\\
    \hline
    \end{tabular}
    \caption{Parameters used for the calculations including longitudinal acoustic phonons, taken from Ref.~\cite{krummheuer2005pure}}
    \label{tab:phonon_parameters}
\end{table}
\section{Experimental Setup}

The experiment is carried out with the sample mounted in an Attodry800 closed cycle system and cooled to base temperature of $T = \SI{6.1}{\kelvin}$. The use of piezo positioners with nanometer precision, combined with focusing the laser beam through an Attocube LT-Apo/NIR objective, enables the targeting of individual quantum dots. The cross-polarization filtering is realized with a high quality polarizing beam splitter (PBS) (B. Halle Nachfolger) and nanoparticle linear film polarizer (Thorlabs/Codixx) in the excitation and detection path. A quarter waveplate (QWP), mounted in a precise motorized rotation stage (PI), is introduced between the PBS and cryostat to minimize elliptical polarization components. Finally coupling the detection path into a single mode fiber serves also as spatial filter to further reject back scattered laser light. 
A high precision and low noise DC voltage source (SRS DC205) connected with coaxial cables and cryo feedthroughs is employed for gating the diode structure.

\begin{figure}[ht]
	\includegraphics[width=\columnwidth]{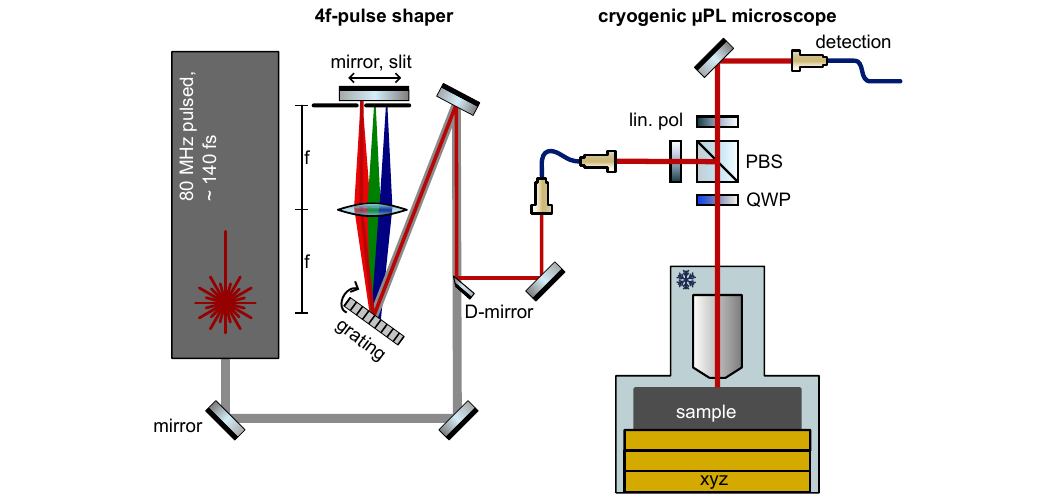}
	\caption{Sketch of the folded 4f-pulse shaper setup used to adjust wavelength and duration of the excitation pulse. The width of the slit determines the duration of the carved out pulse while the position in the Fourier plane selects the center wavelength. The pulses are fiber coupled to a confocal, cryogenic $\mu$PL setup. The back scattered laser light is suppressed in the detection path by cross-polarized filtering, realized with thin-film polarizers, a polarizing beam splitter and a quarter wave plate. In addition, spatial filtering of the detection path is done by fiber coupling into a single mode fiber.}
	\label{fig:setup}
\end{figure} 

The initial laser pulses with duration $\tau = \SI{140}{\femto\second}$ are generated by a passive mode-locked Ti:Sa laser (Coherent Mira900f) and sent free space to a folded 4f pulse shaper setup. Using a blazed grating with 1800 l/mm and a lens with focal length $\SI{300}{\milli\meter}$ carving pulses with up to $\sim\SI{12}{\pico\second}$ duration is possible. An adjustable slit is used for spectral manipulation in the Fourier plane. The shaped pulses are characterized by non-colinear intensity autocorrelation measurements (APE pulseCheck NX) and spectral analysis.

The GaAs quantum dot investigated in this paper was grown by molecular beam epitaxy via local droplet etching. In a first step in-situ deoxidation of a semi insulating GaAs (001) substrate is followed by deposition of a GaAs buffer layer and growth of a AlAs/GaAs superlattice (30 x ($\SI{2.5}{nm}/\SI{2.5}{nm}$)) to planarize the surface and bury impurities on the wafer surface. This is followed by another GaAs buffer layer and a planar distributed Bragg reflector (DBR) composed of 10 pairs Al$_{0.95}$Ga$_{0.05}$As/Al$_{0.15}$Ga$_{0.85}$As. An additional Al$_{0.95}$Ga$_{0.05}$As layer is used to match the anti-node of the electric field with the QD position in the planar $5\lambda /2$ cavity which is formed by including four more pairs on top of the final structure to enhance the extraction efficiency. 
The quantum dot is embedded in a p-i-n diode structure to stabilize the charge environment and allow deterministic charging.
The n-side of the diode, consisting of Al$_{0.15}$Ga$_{0.85}$As doped with Si (n$_\mathrm{Si}$=7.4$\cdot10^{17}$ cm$^{-3}$), serves as an electron reservoir for the QDs. The lower Al concentration of $\SI{15}{\%}$ prevents the formation of deep donor levels (DX centers) \cite{Mooney1990,Munoz1993}.
The tunneling barrier between the n-doped layer and the quantum dot consists of a low temperature $\SI{5}{nm}$ Al$_{0.15}$Ga$_{0.85}$As layer to reduce Si segregation, followed by a higher temperature $\SI{10}{nm}$ Al$_{0.15}$Ga$_{0.85}$As layer, as well as a $\SI{15}{nm}$ Al$_{0.33}$Ga$_{0.67}$As layer.
Al is then evaporated without arsenic background to form droplets on the Al$_{0.33}$Ga$_{0.67}$As layer. Holes are etched into the surface due to the As gradient between droplets and surface, resulting in about $\SI{8}{nm}$ deep nanoholes~\cite{Heyn2011}. They are subsequently filled with GaAs to form QDs and capped with another layer of Al$_{0.33}$Ga$_{0.67}$As. More details can be found in~\cite{daSilva2021}.
The quantum dots are separated by a $\SI{268.4}{nm}$ Al$_{0.33}$Ga$_{0.67}$As layer from the p-contact of the diode, which is grown by carbon doping n$_\mathrm{C}$ = $5\cdot10^{18}$~cm$^{-3}$ a $\SI{65.6}{nm}$ thick Al$_{0.15}$Ga$_{0.85}$As to form a p+ region, followed by a $\SI{13}{nm}$ highly doped ({n$_\mathrm{C}$ = $9\cdot10^{18}$}~cm$^{-3}$) p++ region. 
After the top DBR the structure is capped with $\SI{4}{\nano\meter}$ of GaAs to protect the AlGaAs layers from oxidizing.\\
To contact the n-layer, trenches are etched using sulfuric acid followed by deposition of Ni/AuGe/Ni/Au and an annealing step at $\SI{420}{^{\circ}C}$ for two minutes in forming gas environment. For the p-contacts Pt/Ti/Pt/Au was used (annealed at $\SI{400}{^{\circ}C}$ for two minutes). 

\section{Pulse Shape Dependence}

\begin{figure}[ht]
	\includegraphics[width=\columnwidth]{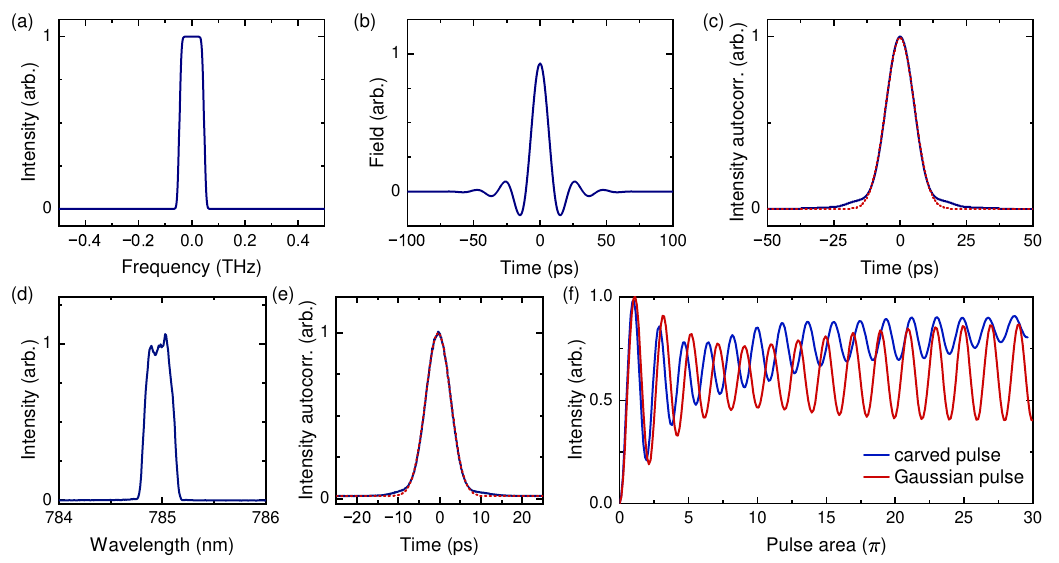}
	\caption{Influence of the carved pulse shape on Rabi rotations (a) Theoretical spectrum of the carved pulse with width $\SI{0.1}{\tera\hertz}$ from an initial Gaussian pulse with $\SI{3}{\tera\hertz}$ FWHM. (b) Corresponding envelope of the electric field of the carved pulse approximating a sinc(x)-function. (c) The resulting intensity autocorrelation function of the carved pulse exhibits additional shoulders compared to a Gaussian pulse shape (red dashed line). (d) Experimental spectrum of a carved $\SI{5}{\pico\second}$ long pulse and the (e) measured intensity autocorrelation function plotted together with a Gaussian fit (red). (f) Theoretical Rabi rotations driven by the carved pulse (blue) show a stronger damping and increasing average with stronger driving compared to a Gaussian pulse.}
	\label{fig:4}
\end{figure}

We employ a self-built 4f pulse shaper, as described in the previous section, to prepare the laser pulse for resonant excitation. Given an initial slightly chirped $\SI{140}{\femto\second}$ long Gaussian pulse with corresponding spectral FWHM of $\SI{7.16}{\nano\meter}$ allows precise adjustment of the wavelength of the carved pulse by simply moving the slit in the Fourier plane. Furthermore, the duration of the carved pulse is manipulated by adjusting the width of the slit. Taking into account the diffraction at the edges of the slit, the resulting spectral shape can be approximated by a steep double sigmoidal function, as depicted in Fig.~\ref{fig:4}(a, theory) and (d, experiment). Consequently, the corresponding electric field amplitude reminds of the sinc-function, as shown in Fig.~\ref{fig:4}(b). This temporal shape results in a Gaussian curve with additional shoulders in the intensity autocorrelation, as presented in Fig.~\ref{fig:4}(c) and (e). The deviation from a pure Gaussian pulse has a significant impact on the Rabi rotation. The simulated Rabi rotations for a Gaussian (red) and carved (blue) $\SI{9.0}{\pico\second}$ long pulse are presented in Fig.~\ref{fig:4}(f). For the carved pulse the oscillation features stronger damping and a rising mean value, which is an obstacle to observe the reappearance of Rabi rotations.  
    
\section{Simulation for different temperatures} \label{sec:app:temp}
\begin{figure}[h]
	\includegraphics[width=0.49\columnwidth]{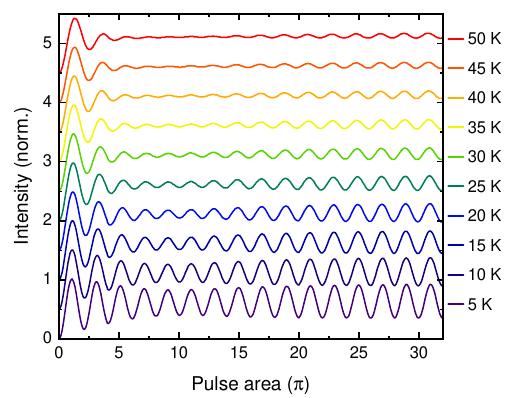}
    \includegraphics[width=0.49\columnwidth]{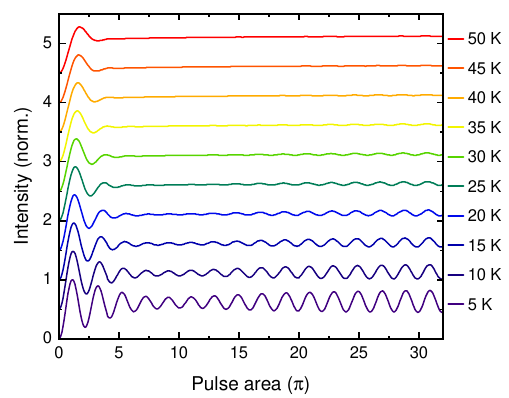}
	\caption{(a) Rabi rotations driven by a \SI{9.0}{\pico\second} long pulse for increasing temperature from $\SI{5}{\kelvin}$ to $\SI{50}{\kelvin}$. For elevated temperatures the overall damping increases while the point of maximum damping remains at $\sim 9\pi$. The reappearance of Rabi rotations is still preserved at $\SI{50}{\kelvin}$. (b) Comparison with 1.5 times stronger phonon coupling.}
	\label{fig:temp_theo}
\end{figure} 

We have performed numerical simulations of the temperature series of the Rabi rotations for a $9.0$~ps long pulse for two different coupling strengths in Fig.~\ref{fig:temp_theo}. While (a) refers to the literature values of the deformation potential coupling strength, in (b) we multiplied the coupling strength by a factor $\lambda=1.5$.

We find the same behavior as described in the main manuscript: Due to the effectively increased phonon coupling strength, for increasing temperatures the damping of the Rabi rotations becomes stronger. Nonetheless, the reappearance is still visible up to high temperatures of $50$~K in (a) und $30$~K in (b). Comparing the theoretical results with the experimental data in Fig.~\ref{fig:temp_exp}, we find that for temperature up to $\SI{30}{\kelvin}$ the data is matched well by the theoretical data shown in (b), while for higher temperatures, the experiments are better described by (a). We attribute this change in phonon coupling strength to the change of the applied voltage to find the resonance condition.

\section{Pulse duration dependence}

\begin{figure}[h]
	\includegraphics[width=\columnwidth]{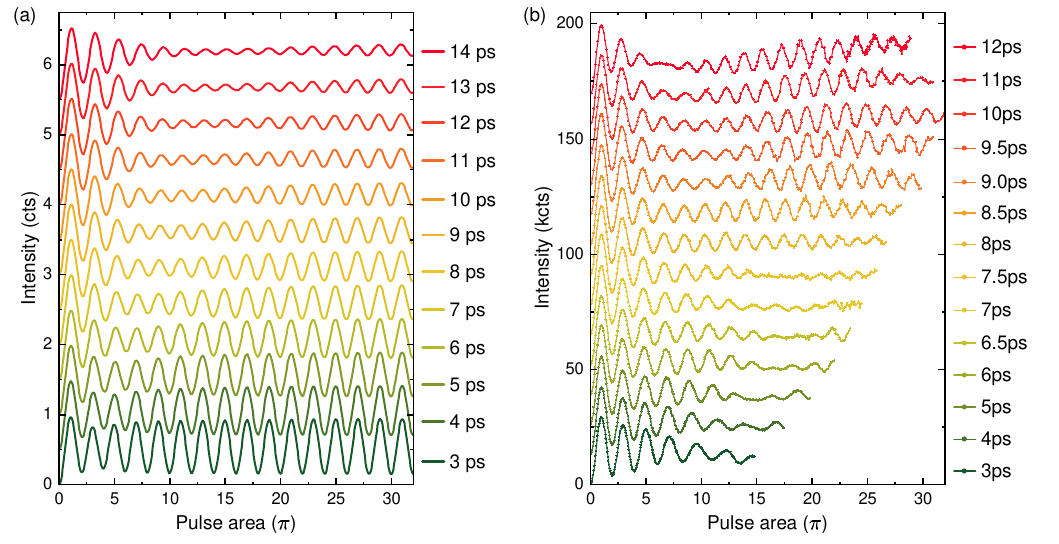}
	\caption{(a) Calculated series of Rabi rotations for varying pulse duration. Longer pulses require larger pulse areas to reach the electric field amplitude to strongly dress the two level system to overcome the phonon spectral density. (b) The experimental data exhibits a similar trend for short pulses.}
	\label{fig:duration}
\end{figure} 

The Rabi rotations depend on the pulse area, which in theory was encoded by $\alpha$. The pulse area can also be derived from the electric field $E(t)$ and the dipole moment $d$ via the integral
\begin{equation}
    \alpha \propto \int_{-\infty}^{\infty}{dt \, d \cdot E(t)}
\end{equation}
For pulsed excitation, the pulse area therefore depends either on the pulse strength (that was varied previously) or the pulse duration.

Therefore, it is of interest to study also the pulse duration dependence. Figure~\ref{fig:duration}(a) shows the numerical simulation of the Rabi rotations for pulse durations from $\SI{3}{\pico\second}$ to $\SI{14}{\pico\second}$. For short pulses the phonon damping increases rapidly with the maximum at $4\pi$, while for longer pulses this point shifts to higher pulse areas with a more pronounced attenuation of the amplitude reflecting the interplay with the phonon spectral density. 

The experimental data, presented in Fig.~\ref{fig:duration}(b) exhibits the same trend for pulse duration up to $\SI{7.5}{\pico\second}$ but a reversed shift for further increasing pulse area. We suspect the altering temporal shape of the laser pulse, caused by the pulse carving to affect the damping as well. Note that the pulse area scales inversely with the duration $\sim 1/\tau$ limiting the experimentally accessible pulse area for shorter pulses.

\section{Detuning}
\begin{figure}[h]
	\includegraphics[width=\columnwidth]{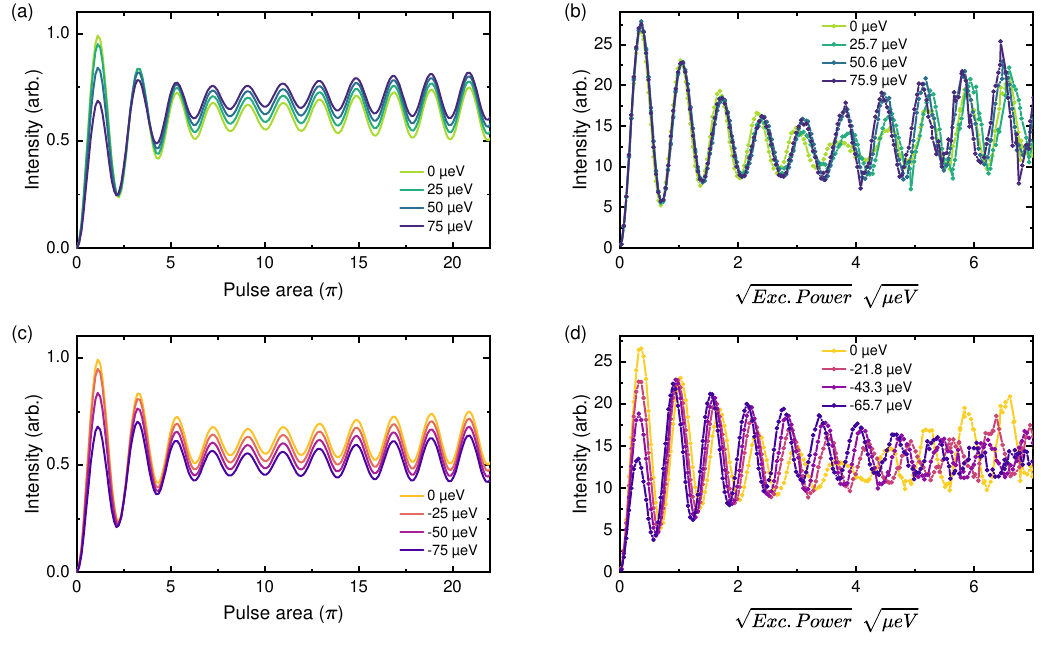}
	\caption{Influence of laser detuning on the Rabi rotations: (a) and (c) show theoretical calculations for positive and negative detuning, respectively. With increasing detuning the inversion of the population for a $\pi$-pulse is reduced while the mean value of the oscillation increases for positive detuning and decreases for a negative detuned laser. (b), (d) Experimental data show good qualitative agreement with the theoretical model.}
	\label{fig:detuning}
\end{figure} 
As a final step, we move away from the resonance condition for the excitation and analyze the Rabi rotations for detuned excitation. For small detunings, still Rabi rotations are expected, but with a less pronounced amplitude. In addition, phonon-assisted processes might be present. 

Given the setup configuration of the 4f-pulse shaper, the tuning rate of the wavelength of the carved pulse is $\SI{2.3}{\micro\eV/\micro\meter}$ for displacement of the slit in the Fourier plane. This makes the detuning of the laser to the driven transition susceptible to slight mechanical drifts. 

In addition, we observe a power dependent blue shift of the emission line which we attribute to the creation of charge carriers in the intrinsic region of the diode structure. To maintain resonant excitation for power-dependent measurements this shift is compensated via a controlled quantum confined Stark effect tuning by dynamically adjusting the gate voltage. However, a temporal dependence as well as non-perfect compensation cannot be fully excluded introducing an additional source for unwanted detuning.

The calculated Rabi rotations for different detunings in steps of $\SI{25}{\micro\eV}$, presented in Fig.~\ref{fig:detuning}(a) and (c), show a reduced emission intensity for $\pi$-pulse excitation which is balanced out for stronger driving by phonon-assisted excitation. This results also in a positive offset of the mean value for blue detuned excitation. The experimental data for positive detunings, shown in Fig.~\ref{fig:detuning}(b) lacks the feature at $A = 1\pi$ but reproduces the increase of the mean value with increasing driving strength. For negative detuning, Fig.~\ref{fig:detuning}(d), we observe a pronounced reduced population inversion for $1\pi$ in agreement with the theoretical model. Moreover, the point of maximum damping of the Rabi rotation is shifted to larger pulse areas with increasing detuning. We attribute part of the deviation from the theoretical model to an asymmetrical pulse spectrum. 

\end{widetext}

\end{document}